\documentstyle[aps,prb,multicol,amsmath,epsfig]{revtex}
\begin{document}
\draft
\title{Charging effects in a quantum wire with leads}

\author{V.~A.~Sablikov\cite{e-mail}}
\address{Institute of Radio Engineering and Electronics,
Russian Academy of Sciences, Fryazino, Moscow District,
141120, Russia}
\author{S.~V.~Polyakov}
\address{Institute for Mathematical Modelling,
Russian Academy of Sciences, Miusskaya
sq.4a, Moscow, 125047, Russia}
\author{M.~B\"uttiker}
\address{D\'epartement de Physique Th\'eorique, Universit\'e de Gen\'eve,
CH-1211, Gen\'eve 4, Switzerland}
\date{published in Phys. Rev. B, {\bf 61}, No 20, 13763-13773 (2000)}
\maketitle

\begin{abstract}
We investigate the distribution of the electron density and the potential in
a quantum wire coupled to reservoirs, treating this structure as a unified
quantum system and taking into account the Coulomb interaction of electrons.
The chemical potential difference that exists between a decoupled, isolated
quantum wire and the reservoirs gives rise to charge transfer in the coupled
system.  We show that the quantum wire can be charged positively or negatively
or remain neutral as a whole, depending on such factors as the wire radius
and the background charge density in the wire.  The magnitude of the charge
and its sign are to a large extent determined by the exchange interaction of
the electrons in the wire. Using a Hartree-Fock approach, we develop a model
of a quantum wire which includes the reservoirs. This model allows us to find
the self-consistent distribution of the electron density and the potential in
the wire both at equilibrium and in the presence of transport.  The linear
conductance is investigated as a function of the chemical potential.  The
nonadiabatic transition from the reservoirs to the wire leads to conductance
oscillations caused by multiple scattering of electron waves.  The period of
the oscillations depends on the charge acquired by the wire and the exchange
energy.  We find that the exchange interaction strongly enhances the Friedel
oscillations near the contacts.  However, they do not noticeably suppress the
conductance because the wire has a finite length and is charged.  Under far
from equilibrium conditions, which appear when the applied voltage exceeds the
Fermi energy in the wire, the system becomes unstable with respect to
fluctuations of the electric potential and the electron density. The
instability results in the appearance of multistable electron states.
%
\end{abstract}

\vspace{1cm}
\begin{multicols}{2}

\section{Introduction}
Electron transport in quantum wires (QWs) now attracts considerable
interest because of the fundamental importance of the electron-electron
({\em e-e}) interaction in one-dimensional (1D) systems.~\cite{Voit,Schulz}
In addition, it is expected that the {\em e-e} interaction may produce
important effects in the transport that would be attractive for applications.
However, up to now there is no clear and unambiguous knowlege of which effects
of {\em e-e} interaction are observable under the realistic conditions of
a QW with leads and what the reasons are for the deviations from conductance
quantization (observed experimentally~\cite{Thomas}), and very little is
known about the electron transport under far from equilibrium conditions.

In order to understand the transport properties, it is necessary to know the
spatial distribution of the electric potential and the electron density in
the structures under investigation. Many arguments and
facts~\cite{Maslov,Ponomarenko,Safi,Yacoby,Cheianov} show that the leads play
an essential role in the conductance, if the transport is investigated by
measuring the electric current in an external circuit and by measuring the
voltage drop between the leads.  It is obvious that the interaction
of the QW with the leads is not weak.

In the present paper we investigate this interaction, considering the QW and
the leads as a unified system. An important problem that arises is how the
electric field and the electron density are distributed at equilibrium (i.e.,
in the ground state), and how they are redistributed when an external
voltage is applied. We show that even in the equilibrium state the QW
acquires a charge and a contact potential difference between
the QW and the leads. This phenomenon is similar to the well known contact
potential difference in classical conductor systems.  The essential
difference between the low-dimensional system under consideration and
classical conventional 3D conductors is that the contact potential is not
screened over a finite length but spreads over a length determined by the
geometrical size of the structure.~\cite{Odintsov} The contact potential
difference is determined by the difference in the chemical potentials $\Delta
\mu$ of electrons in the QW and in the leads, when these subsystems are
considered independently. By analyzing the chemical potentials we find
that three cases are possible, depending on the wire radius and the background
charge density:  (i) the QW is charged positively, (ii) the QW
acquires a negative charge, or (iii) the QW remains neutral as a whole.

We develop a model of a QW with leads in which the lead-wire interaction
is taken into account. It is based on a Hartree-Fock approach for the
electrons in the QW and the representation of leads (which are considered as
electron reservoirs) in a way that takes into account their 3D nature, but
requires only a 1D calculation.  Using this model we investigate the
distribution of the electron density and the electric potential at
equilibrium, as well as under far from equilibrium conditions.  In
particular, we emphasize the role of the exchange interaction effect on these
quantities.  If in the absence of coupling the chemical potential of the wire
exceeds that of the reservoir, $\Delta\mu>0$, the coupled system expels
electrons from the QW into the leads, when equilibrium is established. As
a result a potential well appears in the QW.  The wire acts as a charge donor
to the reservoir.  If $\Delta\mu<0$, electrons are attracted by the
wire, and a potential barrier arises between the QW and the leads. The wire
acts in this case as an acceptor.  In the case where $\Delta\mu=0$, the
electron density is redistributed only within the QW and only Friedel
oscillations arise near the contacts.  The Friedel oscillations appear in all
cases; however, they are superimposed on a much more slowly varying potential
created by the acquired charge. The slowly varying potential has an amplitude
that is large compared to the Friedel oscillations and thus dominates the
scattering processes.  The exchange interaction strongly affects the
potential shape and somewhat enhances the Friedel oscillation amplitude.

When an external voltage is applied, the chemical potentials in the electron
reservoirs are shifted relative to each other, disturbing the electron flows
in the QW. As this takes place, the electron density, the potential, and the
exchange energy are changed self-consistently. The importance of the
electrostatic potential distribution in quantum wires with leads, especially
for the investigation of time-dependent transport, and for nonlinear
transport, has been emphasized previously,~\cite{Buttiker} but quantitative
calculations have to our knowledge not been reported thus far.

Under far from equilibrium conditions, when the applied voltage exceeds
the Fermi energy in the QW, the electron density is substantially
redistributed between the QW and the reservoirs, giving rise to a very strong
variation in the potential landscape. In turn the potential produces a
variation in the electron density. The connection between the electron
density and the potential is very important for the understanding of
nonlinear transport. The need of a self-consistent treatment has been
emphasized by Landauer.~\cite{Landauer} Within the scattering approach, in
the weakly nonlinear regime, it has been investigated by Christen and one of
the present authors \cite{Christen} and by Ma, Wang and Guo.~\cite{Guo} The
calculation of the present paper allows us to investigate the strongly
nonlinear transport in the system under consideration.  If the applied
voltage is high enough, the self-consistent connection of charge and
potential gives rise to an instability of the electron density distribution
and ultimately leads to multistability of the electron states in the QW.
This means that several stable states with different spatial distributions of
the electron density and the potential are possible at a given applied
voltage.

The paper is organized as follows. In Sec.~\ref{Contact-potential} the
chemical potential difference between the decoupled QW and the 2D electron
reservoir is analyzed. Section~\ref{model} describes the model of the QW with
leads.  Section~\ref{eq-state} contains the results of the numerical
calculations of the electron density and potential distribution in the QW
with leads. In Sec.~\ref{multi-stability} the multistability of the electron
states is described that appears for far from equilibrium conditions.

\section{Contact potential difference}
\label{Contact-potential}
To be specific we consider a QW connecting two regions of a 2D electron gas.
We assume that there are no nearby gates and that all electric field
lines emanate and terminate either on the wire or on the 2D electron gas.
QW structures of this kind are produced by etching of heterostructures with a
2D electron gas. Such structures are widely used in
experiments.~\cite{Tarucha,Ploner,Wesstrom,Maximov,Novoselov}

First we investigate such a QW separately from 2D electron reservoirs. The
uncoupled wire is charge neutral. The electron charge is concentrated inside
the QW, while the compensating positive charge of the impurities is really
located in the immediate vicinity of the QW or at its surface. The decoupled
QW and the electron reservoir have their own chemical potentials
$\mu_{\rm 1D}$ and $\mu_{\rm 2D}$ which are generally not equal each other. We are interested in
the chemical potential difference $\Delta \mu$ between the QW and the
reservoir.

According to Seitz's theorem~\cite{PinesNozieres} the
chemical potential in the QW is determined by the Fermi energy of
noninteracting electrons and the self-energy $\Sigma (k_F)$ which takes into
account the {\em e-e} interaction,
\begin{equation}
\label{mu1D}
\mu_{\rm 1D} = \varepsilon_0 + \varepsilon_F + \Sigma(k_F) \,.
\end{equation}
Here $\varepsilon_0\approx \pi^2\hbar^2/(2ma^2)$ is the first subband energy
caused by transverse confinement ($a$ is the QW radius, $m$ is the effective
mass of electrons). $\Sigma(k_F)$ contains the contributions arising from the
exchange and correlation interaction as well as from the electron interaction
with the positive background charge. The exchange and correlation energy was
investigated in the recent paper of Calmels and Gold~\cite{Calmels} using the
self-consistent theory of Singwi, Tosi, Land and Sj\"olander~\cite{STLS} for
the case where only the lowest subband is occupied. The Hartree energy is
easily estimated if we assume that the positive charge is located at the
surface of the QW.  These calculations lead to the following expression for
the chemical potential $\mu_{\rm 1D}$ of the QW in terms of the dimensionless
parameters $r_s = 1/(2a_Bn)$ and $\beta=a_B/a$ (with $a_B$ the
effective Bohr radius and $n$ the 1D electron density):
\begin{equation}
\label{mu1d}
\frac{\mu_{\rm 1D}}{R_y} \approx \frac{\pi^2}{2}\beta^2 +
\frac{\pi^2}{16 r_s^2} + \frac{\Sigma_{\rm xc}}{R_y} -
\frac{B_H}{r_s}\,,
\end{equation}
where $R_y$ is the effective Rydberg. In Eq.~(\ref{mu1d}) the first term
is the lowest subband energy, the second term is the kinetic energy, and
the third term represents the exchange and correlation energy. There are two
expressions for the exchange-correlation term depending on whether $r_s<1$ or
$r_s>1$:
$$
\frac{\Sigma_{\rm xc} }{R_y} \approx
- \beta \frac{5.57\pi - 4 \beta r_s}{2\pi^2}\,\qquad  \mbox{if}\quad r_s<1\,
$$
and
$$
\frac{\Sigma_{\rm xc} }{R_y} \approx
- \frac{1.84}{r_s}\left[\ln \frac{2 \beta r_s}{\pi} + 0.7115\right]\,
\qquad \mbox{if}\quad r_s>1\,.
$$
The last term in Eq.~(\ref{mu1d}) is the Hartree energy of the electron
interaction with the positive background; $B_H$ is a numerical factor
that depends on the radial distribution of the electron density. If the
electron density is distributed uniformly, $B_H\approx1/3$.

The chemical potential in the 2D reservoir can easily be obtained from the
known expression~\cite{Isihara} for the electron energy as a function of the
density parameter $R_s = (\pi a^2_BN_{\rm 2D})^{-1}$ (where $N_{\rm 2D}$ is
the 2D electron density). In the high density case, $R_s<\sqrt{2}$, one obtains
\begin{equation}
\label{mu2d}
\begin{array}{rl}
\dfrac{\mu_{\rm 2D}}{R_{y}} = & \left(\dfrac{\pi a_B}{d}\right)^2 + 2 R_s^{-2} -
1.80 R_s^{-1} - 0.38 -\\
& \qquad \qquad 0.0863 R_s\ln R_s + 0.519 R_s\,,\\
\end{array}
\end{equation}
where $d$ is the thickness of the 2D layer.

\begin{figure}[htb]
\hspace{-0.5cm}
\mbox{\epsfig{file=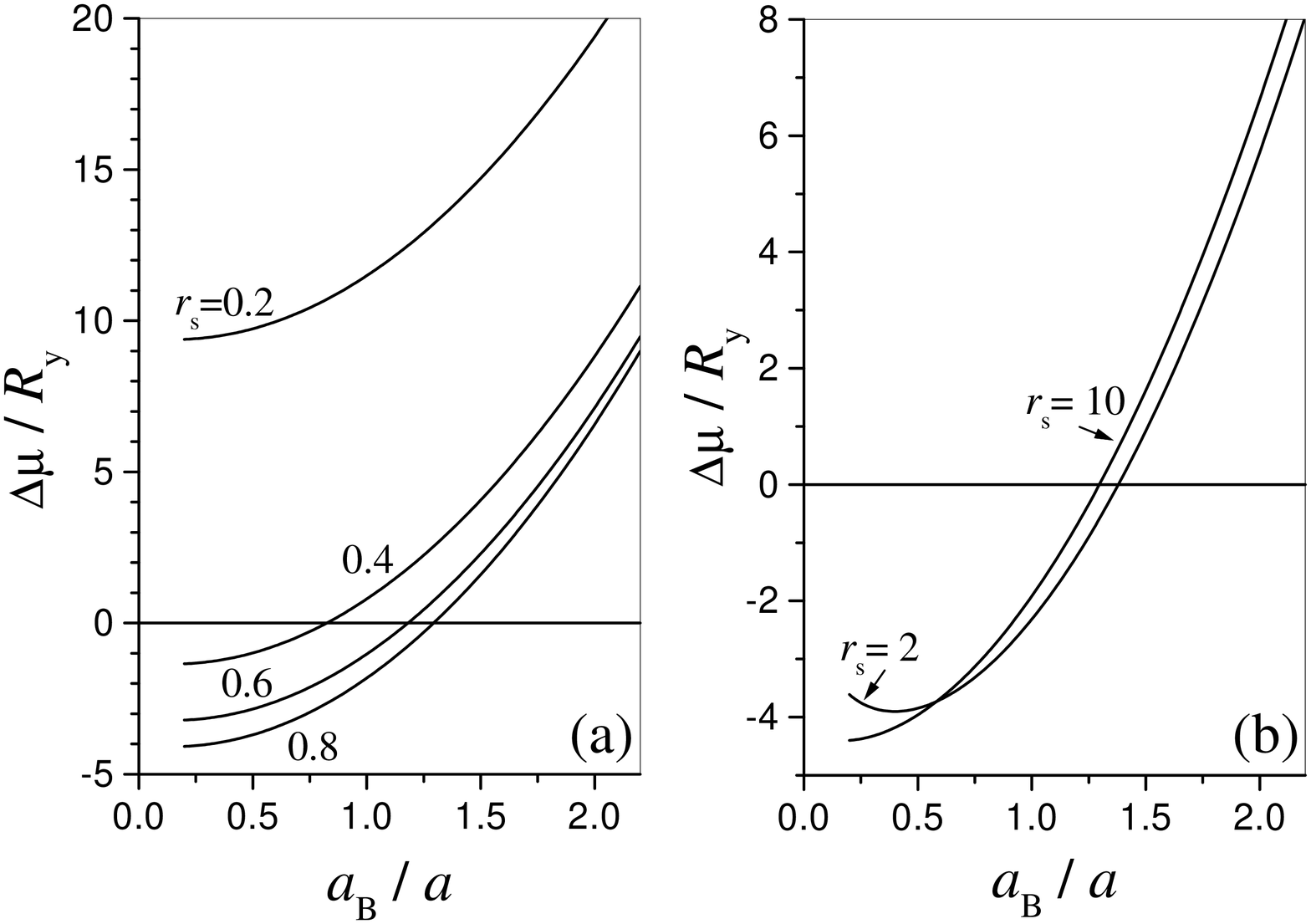,width=9cm}}
\narrowtext
\caption[to]{Dependence of the chemical potential difference $\Delta\mu$
between the decoupled QW and the 2D electron reservoir on the wire radius
$a$ (in units of the Bohr radius) for various density parameters: (a)
$r_s<1$, (b) $r_s>1$. The calculations were done with the following
parameters of the 2D layer: $d=2a$, $R_s=0.5$.}
\label{delta-mu}
\end{figure}

Using Eqs.~(\ref{mu1d}) and (\ref{mu2d}), the chemical potential difference
$\Delta \mu = \mu_{\rm 1D} - \mu_{\rm 2D}$ is calculated as a function of the QW
radius and the density parameter $r_s$. The results of these calculations
are illustrated in Fig.~\ref{delta-mu}, where the $\Delta\mu$ dependence on
the wire radius $a$ is shown for various values of $r_s$. Here we consider
$r_s$ as an independent parameter because the background charge density
depends on external factors, such as the charge absorbed at the wire surface.
Figure~\ref{delta-mu} shows that in sufficiently thin wires the chemical
potential is higher than in the reservoir. However, with increasing radius
of the QW the chemical potential in the wire can become lower than in the
reservoir.

The chemical potential difference $\Delta \mu$ is caused by all energy
components contributing to the chemical potentials in the QW and in the
reservoir. As an example it is instructive to consider the estimations for
the specific case where the QW diameter is equal to the 2D layer thickness,
$d=2a$, and the background charge density per unit area, $N_{\rm 2D}$, is the same
in the reservoir and in the QW. The latter means that $n=2aN_{\rm 2D}$. In this
case the energies contributing to the chemical potentials are estimated as
follows.  The confinement energy in the QW is approximately twice that in
the reservoir. Hence the confinement energy causes $\mu_{\rm 1D}$ to rise with
respect to $\mu_{\rm 2D}$. The ratio of kinetic energy in the QW to that in the
reservoir is $\varepsilon_{F1}/\varepsilon_{F2}\sim an$. Since the QW is
supposed to be a 1D system, the product $an$ must be small. Hence the Fermi
energy $\varepsilon_{F1}$ in the QW is noticeably smaller than the Fermi
energy $\varepsilon_{F2}$ in the reservoir. This results in lowering
$\mu_{\rm 1D}$ relative to $\mu_{\rm 2D}$. The ratio of the
exchange-correlation energies in the case of $r_s<1$ is estimated as
$\varepsilon_{\rm xc1}/\varepsilon_{\rm xc2}\sim (an)^{-1/2}$. The
exchange-correlation energy in the QW is seen to be larger than that in the
reservoir. Taking into account that the exchange-correlation energy is
negative, we conclude that it lowers $\mu_{\rm 1D}$ with respect to
$\mu_{\rm 2D}$.

If the wire is now coupled to the 2D reservoir, we can distinguish three
cases.  For $\Delta \mu > 0$, electrons are transferred from the QW to the
reservoir, so that the positive background charge dominates the electron
charge in the wire. If $\Delta \mu < 0$, the reservoir supplies electrons
to the QW producing an excess negative charge. When $\Delta \mu= 0$, the
electron density is not redistributed between the wire and
reservoir.~\cite{footnote1}

The electron density redistribution continues until an equilibrium state
is attained in the whole system such that there exists a uniform
electrochemical potential. As this takes place, a charge and a built-in
electric field $E_{0}$ appear in the QW. If the reservoir conductivity is
high, the built-in field satisfies the condition $e\int_0^{\infty}dx\, E_0(x) =
\Delta \mu$. Such a charge transfer is similar to what happens when a contact
potential difference appears in classical 3D systems. However, an essential
difference is in the distance over which the contact field is screened. In
the 3D case the contact field is screened over a finite length (e.g.,
the Debye length or the Thomas-Fermi screening length).  In the mesoscopic
structure considered here, the contact field is produced by the charge, one
part of which is situated in the 1D wire and the other on the
surface of the reservoir adjacent to the wire.  One can easily see, that owing
to the 3D nature of the electric field, any distribution of charges in a QW
cannot screen the contact field over a finite distance.  Thus the question
arises of how the charge density and the electric field are distributed in the
QW and over the reservoir surface. In the case of a high enough electron
density, the interaction effects are not strong, and this problem can be solved
analytically using the Thomas-Fermi approximation.~\cite{Sablikov3} In the
present paper, we study this problem using a numerical solution of the
Schr\"odinger equation coupled to the Poisson equation within the
Hartree-Fock approximation.

\section{Model}
\label{model}
Finding the self-consistent electron density and the potential in a QW
coupled to electron reservoirs is a rather complicated problem because the
electron density redistribution between the QW and the reservoirs produces a
strong variation of the electron energy in the QW. The energy variation is
estimated as $\Delta \mu$, which is shown above to be of the order of the
Fermi energy. In essence, the QW and the reservoirs should be considered as a
unique quantum system. To our knowledge, such a problem has not been studied to
date. In the present paper, we investigate it using a simplified model based
on the Hartree-Fock approximation, which allows one to take adequately into
account the charge accumulated in the QW both for the equilibrium state and
under the far from equilibrium conditions appearing when an external voltage
is applied. This approach allows one also to study the exchange interaction
effect on the charge accumulated and the electric potential. However, it does
not take into account the electron correlation energy. The ratio of the
correlation energy to the exchange energy depends on the electron density. For
1D conductors this ratio can be estimated according to
Ref.~\onlinecite{Calmels}. The correlation energy is negligible when
$r_s\le 1$. Under this condition the Hartree-Fock approximation is
justified.

It is instructive to estimate numerically the number of electrons in a
GaAs QW when $r_s=0.5$. The Fermi energy and the electron density are
respectively $\varepsilon_F\approx$~4~meV and $n\approx 6\times
10^{5}$~cm$^{-1}$. Our computation procedure works well when the QW length is
not too large in comparison with the Fermi wavelength. If one puts the QW
length equal to $0.3~\mu$m, the total number of electrons in the QW is
estimated as about 18. Thus the system contains about ten electrons in a QW
open to reservoirs.  The number of electrons really existing in the
wire and their density distribution is determined by the QW length, the
background charge, the applied voltage, and the {\em e-e} interaction energy.
The parameters of the QW estimated above are easily realizable in experiment.

In order to investigate cases with different relative positions of the
chemical potentials in the QW and reservoirs, we introduce a positive
background charge density $e n_b$ in the QW, which is considered as a
parameter of our model. By varying $e n_b$ it is possible to realize any
relative position of the chemical potentials of the uncoupled system. Charges
on the QW surface are not taken into account in our present consideration.

Another simplification is that only the lowest subband in the QW is
considered.

\begin{figure}[htb]
\mbox{\epsfig{file=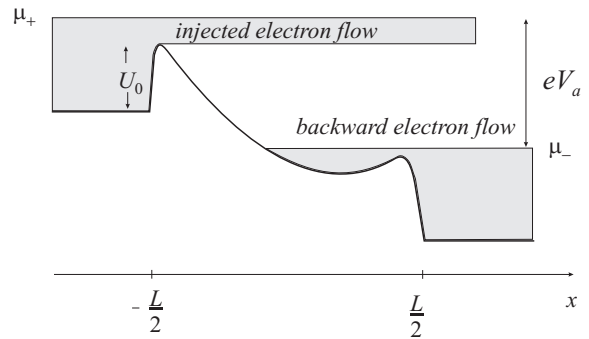,width=8cm}}
\narrowtext
\caption[to]{Energy diagram of a QW with leads and the electron flows.}
\label{diagram}
\end{figure}

A 1D sketch of the energy diagram of a QW with leads and the electron flows
(in a far from equilibrium situation) is shown in Fig.~\ref{diagram}.  Here
$U_{0}$ is the confinement energy in the QW and ${\bf \mu_{\pm}}$ are the chemical
potentials in the reservoirs.  The positions of ${\bf \mu_{+}}$ and ${\bf
\mu_{-}}$ relative to the conduction band bottom of the corresponding
reservoirs are fixed because the electron gas in the reservoirs is
incompressible. However,  ${\bf \mu_{+}}$ and ${\bf \mu_{-}}$ are shifted
relative to each other in the presence of an applied voltage $V_{a}$.

The following energies contribute to the potential shape of the structure:
the confinement energy in the QW; the potential produced by the external
voltage source; the Hartree and exchange energies in the QW.  In the reservoirs
($|x|>L/2$, $L$ being the wire length) the electron density is taken to be so
high that the {\em e-e} interaction energy can be ignored in comparison with the
kinetic energy. Moreover, the reservoirs are assumed to be ideally conducting
and hence can be treated as equipotentials.  Inside the QW the {\em e-e}
interaction potential $U(x,x')$ is determined by both the direct interaction
of electrons with each other and the indirect interaction via image charges
induced by electrons in the reservoirs. Because of this, the interaction
potential depends on the coordinates $x,x'$ of the interacting electrons
separately, rather than simply on their difference.  In the QW the
one-electron wave functions $\psi_{r,k_r,s}(x)$ are characterized by quantum
numbers: $r= \pm 1$ ($r= +1$ indicates electrons incident on the QW from the
left reservoir and $r= -1$ corresponds to electrons incident on the QW from
the right reservoir); $k_r$ is a wave number in the left ($r= +1$) or right
($r= -1$) reservoir; $s$ is a spin variable. In this paper we ignore changes
in $s$ and suppose that the states with opposite spins are equally occupied.
The effects of possible spin polarization in a QW will be considered elsewhere.

The electron transport in the reservoirs, close to the transition between the
reservoir and the QW, is modeled by a 1D Schr\"odinger equation without
interaction. This model allows one to simulate adequately the transmission
probability between the reservoirs and the QW. The value of the transmission
probability calculated in this way as compared with that for a true 2D to 1D
transition at the interface of the reservoir and the wire has been
investigated by direct comparison of the transmission probabilities of these
two geometries.~\cite{Sablikov4} The difference depends on the energy but
does not exceed 15\% even close to the transmission threshold. The physical
reason for the success of the purely 1D model is that near the chemical
potential level, of all the electron waves in the reservoir only those couple
effectively to the wire that have a wave vector that is nearly parallel to
the wire axis (collimation effect~\cite{BH}). We also emphasize that in the
problem under investigation the electron density of the 1D Schr\"odinger
problem is needed only within the wire. It is unimportant for the reservoirs
because they are treated as equipotentials and therefore the calculated
electron density distribution in the reservoirs does not directly affect the
{\em e-e} interaction in the QW. The calculated electron wave functions in the
reservoirs affect only the transmission probability through the contacts.

In the reservoirs, $\psi_{r,k_r,s}(x)$ is thus
\end{multicols}
\widetext
\vspace{-3mm}\noindent\underline{\hspace{87mm}}
\begin{equation}
\label{reservoirWavefunction}
\psi_{r,k_r,s}(x) = \left\{
\begin{array}{ll}
\exp[ik_r (rx+L/2)] + R_r \exp[-ik_r (rx+L/2)]\, &\quad
\mbox{if}\quad r x<-L/2\\
T_r \exp[ik'_r (rx-L/2)] \, &\quad \mbox{if}\quad r x>L/2\,,\\
\end{array}
\right .
\end{equation}
where $(k'_r)^2 = k^2_r + r\times 2meV_a/\hbar^2$, and $V_a$ is the applied
voltage.

In the QW, $\psi_{r,k_r,s}(x)$ is determined by the equation
\begin{equation}
\label{Schrodinger}
-\frac{\hbar^2}{2 m}\,\frac{d^2 \psi_{r,k_r,s}}{dx^2} +
\left[U_0(x) - U_{\rm ext}(x) + U_H(x) + \hat{H}_{\rm ex}\right] \psi_{r,k_r,s} =
\varepsilon_r(k_r) \psi_{r,k_r,s}\,.
\end{equation}
In Eq.~(\ref{Schrodinger}) the potential energy has the following
components.
$U_0(x)$ is an effective potential that simulates the electron confinement
in the QW. In the simplest case, we can assume that
$U_0(x)= U_0 = {\rm const}$ for $|x|<L/2$ and $U_0(x)= 0$ for $|x|>L/2$.
$U_H$ is the Hartree energy,
\begin{equation}
\label{Hartree}
U_H(x) = \int\limits_{-L/2}^{L/2}dx' U(x,x') \left[n(x') -n_b\right] \,,
\end{equation}
with $n_b$ being the positively charged background density, $n(x)$ being the
electron density,
\begin{equation}
\label{e-density}
n(x) = \sum_{r= \pm,s} \int\limits_0^{\infty}\frac{dk_r}{2\pi}
f(k_r)|\psi_{r,k_r,s}|^2 \,,
\end{equation}
and $f(k_r)$ the electron distribution function in the reservoirs. In
Eq.~(\ref{Hartree}) we assume for simplicity that the radial component of the
background charge density is the same as the electron density.

The {\em e-e} interaction potential $U(x,x')$ that appears in Eq.~(\ref{Hartree})
depends on the spatial configuration of the leads. In what follows the
numerical calculations are carried out for the case where the leads are
represented as two plates perpendicular to the QW.  This configuration is
convenient for further calculations because in this case a relatively simple
analytical expression is obtained for $U(x,x')$.~\cite{Sablikov1,Sablikov2}
This form of the interaction potential allows one to take into account not
only the direct Coulomb interaction of electrons but also their interaction
via image charges induced on the lead surfaces. The interaction potential is
$$
U(x,x') = \frac{e^2}{\epsilon L}\int\limits_0^{\infty}
\frac{dy}{\sinh y}|\chi_y|^2\:\left \{
\begin{array}{rcl}
\sinh\left[y\left(1/2 + \xi \right)\right]\,
\sinh\left[y\left(1/2 - \xi '\right)\right] &
{\rm if}& \xi <\xi '\\
\sinh\left[y\left(1/2 - \xi \right)\right]\,
\sinh\left[y\left(1/2 + \xi ' \right)\right]\, &
{\rm if}& \xi >\xi '\,,\\
\end{array}
\right.
$$
where $\xi= x/L$ and $\chi_y$ is the Fourier transform of the radial density,
which is taken to be $\chi_y = \exp \left[-(ay/2L)^2\right]$.
Using the analytical expression for $U(x,x')$, instead of direct solution of
the 3D Poisson equation, highly facilitates computations.

$\hat{H}_{\rm ex}$ is the exchange energy operator,
$$
\hat{H}_{\rm ex}(x) \psi_{r,k_r,s} = \int\limits_{-L/2}^{L/2}dx' U(x,x')
n_{\rm ex}(x,x') \psi_{r,k_r,s}(x')\,,
$$

\noindent\hspace{92mm}\underline{\hspace{87mm}}\vspace{-3mm}
\begin{multicols}{2}
\noindent
where
$$
n_{\rm ex}(x,x') = \sum_{r= \pm}\int\limits_0^{\infty}\frac{dk_r}{2\pi}\left[
\psi^*_{r, k_r,s}(x') \psi_{r, k_r,s}(x)\right] f(k_r)\,.
$$
For the reservoir configuration that we consider here, the "external
potential" is a linear function of $x$:  $U_{\rm ext} = e V_a (x/L+1/2)$.

The energy $\varepsilon_r(k_r)$ in Eq.~(\ref{Schrodinger}) is expressed in
terms of the wave vector $k_r$ and the applied voltage $V_a$,
$$
\varepsilon_r(k_r) = \frac{\hbar^2k_r^2}{2m} + e V_a \delta_{r,-1} \,,
$$
where we assume that the energy reference is fixed at $x=-\infty$.

In addition, we require continuity of the wave functions
determined by Eqs.~(\ref{Schrodinger}) and (\ref{reservoirWavefunction}) and
their derivatives at the reservoir-wire interfaces $x= \pm L/2$.
The distribution functions $f(k_r)$ in the reservoirs are taken in the form
of the Fermi functions with the temperature $T$ considering the fact that the
Fermi level in the right reservoir is shifted down by $eV_a$ with respect to
the left one.

Inside the QW no distribution functions are assigned. The electron
distribution over the energy is determined by the electron flows from the
left and right reservoirs and the interaction processes inside the QW. The
external voltage produces a variation of the electron flows, as a consequence
of which the electron density is changed. Ultimately, this results in the
self-consistent variation of both the electron states and their occupation for
both the left and right moving particles.

The wave functions that we consider in this work are characterized
by a continuous quantum number $k_r$.~\cite{footnote2} Hence, $\psi$ should
be considered as a function of two variables $x$ and $k_r$.
Equation~(\ref{Schrodinger}) is an integro-differential equation with respect
to the variable $x$ and an integral equation with respect to the variable
$k_r$. We develop a numerical scheme for the solution of this equation on a
grid spanning the two variables. The computation method is described in the
Appendix. It is worth noting that in the case where the voltage is applied,
the wave functions are found without using any expansion in terms of the
undisturbed wave functions.

The numerical computations were performed using the 32-processor computer
system Parsytec CC.

\section{The equilibrium state}
\label{eq-state}
First, we consider the equilibrium state that appears in the absence of an
applied voltage $V_a = 0$. In order to realize the three cases ($\Delta\mu >
0,\, \Delta\mu < 0$, and $\Delta\mu =0 $) described in
Sec.~\ref{Contact-potential}, we vary the density of the positive background
charge $n_b$. In doing this it is convenient to compare $n_b$ with the
characteristic density
\begin{equation}
\label{n_0}
n_0 = \frac{2}{\pi \hbar}\sqrt{2m(\mu_0-U_0)}\,,
\end{equation}
where $\mu_0$ is the equilibrium level of the chemical potential in the
system. This quantity has a simple physical meaning in the case where the
exchange and correlation interaction is absent. It is the background density
that determines which of the three cases is realized in the Hartree case. If
$n_b= n_0$, electrons are not redistributed between the QW and the leads in
the equilibration process, if $n_b>n_0$, electrons flow from the QW to the
reservoirs, and if $n_b<n_0$, electrons are transferred from the reservoir to
the QW. Of course, turning on the exchange and correlation interaction shifts
the value of the background density at which electrons are not redistributed.
Nevertheless, as a reference, the value $n_0$ remains convenient.

\begin{figure}[htb]
\mbox{\epsfig{file=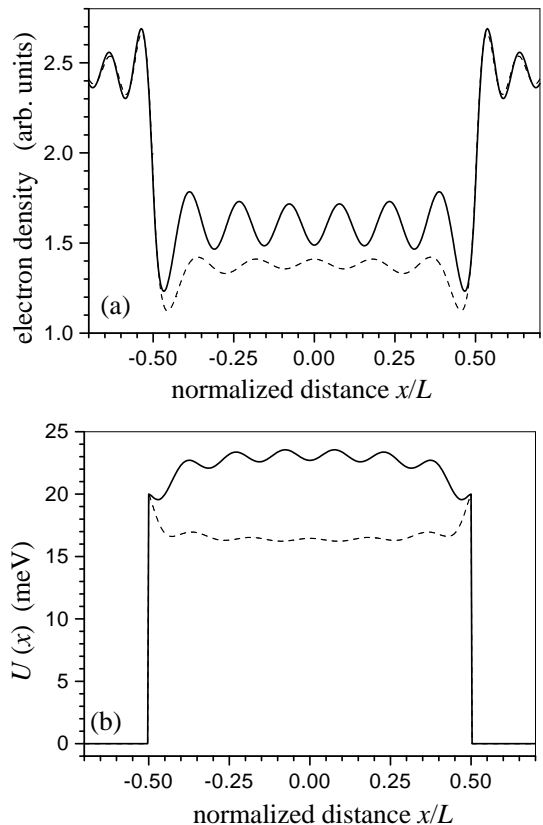,width=8cm}}
\narrowtext
\caption[to] {(a) The electron density distribution and (b) the potential
energy shape in the QW for $n_b = 1.5 n_0$. The dashed lines represent the
case without exchange interaction. The solid lines are obtained by taking the
exchange interaction into account. The calculations were done for the
parameters:  $a= 5\times 10^{-7}$~cm, $L/a= 30$, $U_0$ = 20~meV, $\mu_0-U_0 =
4$~meV, $T$=0, and $V_a$=0.}
\label{n-b15}
\end{figure}

\begin{figure}[htb]
\mbox{\epsfig{file= 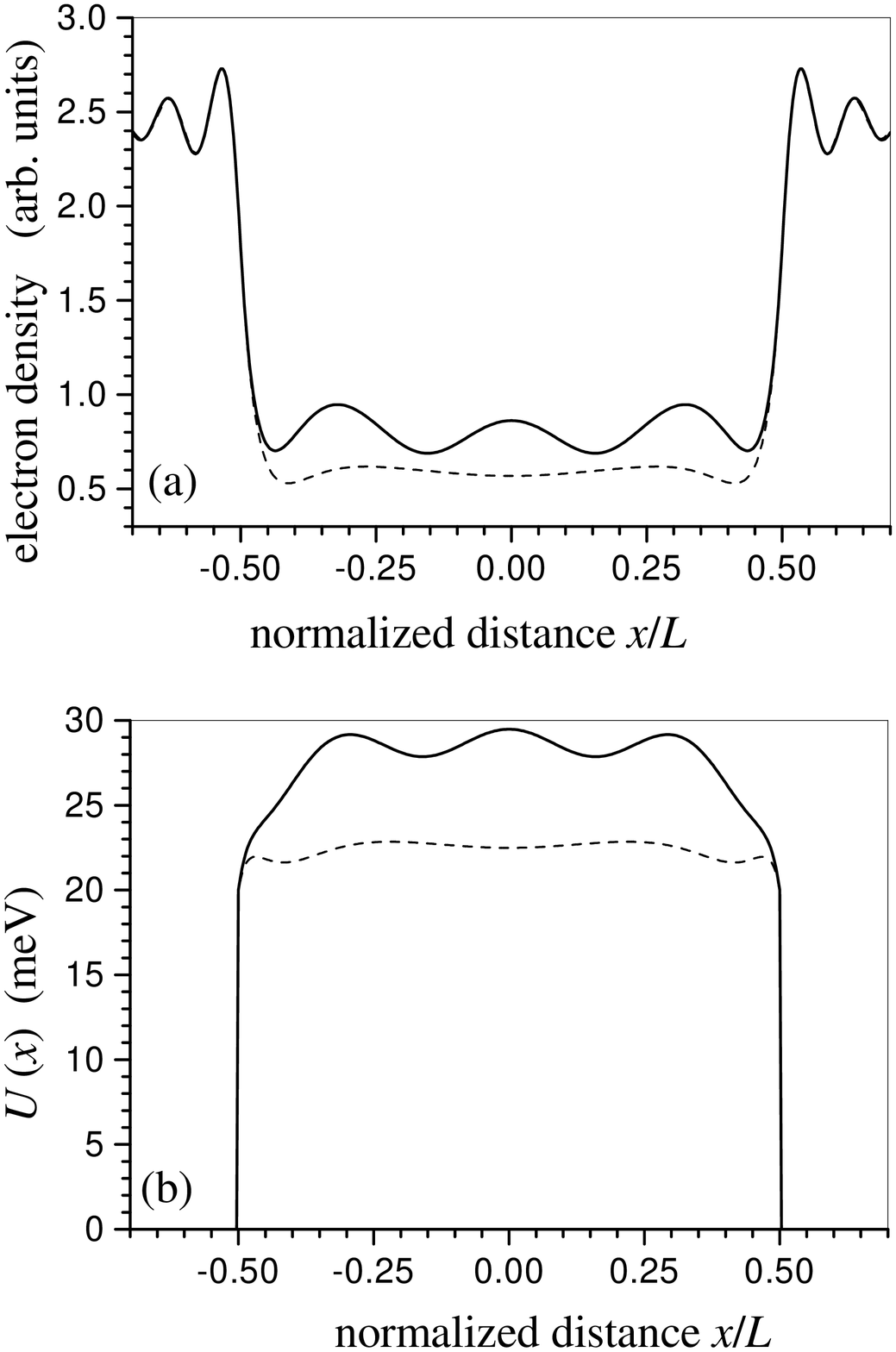,width= 8cm}}
\narrowtext
\caption[to]{The same as in Fig.~\ref{n-b15} but for $n_b= 0.5 n_0$.}
\label{n-b05}
\end{figure}

\begin{figure}[htb]
\mbox{\epsfig{file=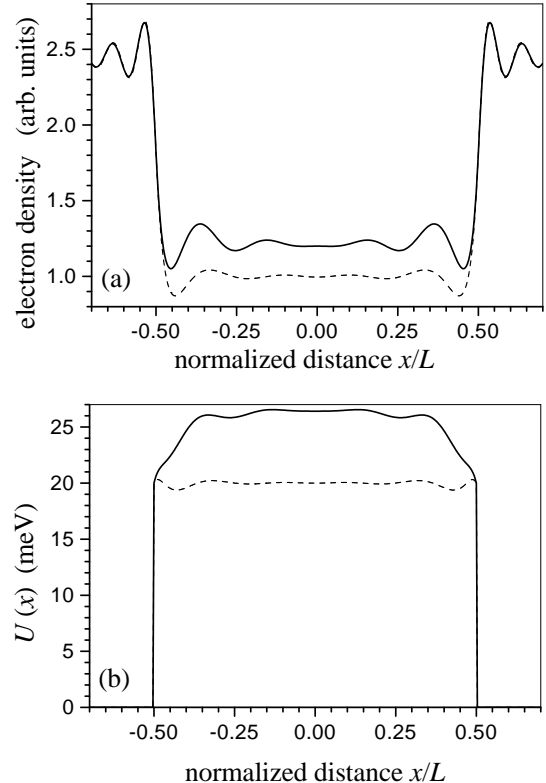,width=8cm}}
\narrowtext
\caption[to]{The same as in Fig.~\ref{n-b15} but for $n_b=1.0 n_0$.}
\label{n-b10}
\end{figure}

The electron density distribution $n(x)$ for the three cases is illustrated
in Figs.~\ref{n-b15}, \ref{n-b05}, and \ref{n-b10}.  Shown here are also the
graphs of the potential energy $U(x)$ that includes the confinement energy
and the Hartree energy,
$$
U(x) = U_0 + U_H\,
$$
but not the exchange energy. The exchange energy is not included in $U(x)$
because it is a functional of $\psi$ rather than a direct function of $x$.
It is useful to note that $U_H$ essentially coincides with the electric
potential.

Let us consider first the case where the exchange interaction is not taken
into account. It is illustrated by the dashed lines in Figs.~\ref{n-b15},
\ref{n-b05}, \ref{n-b10}. In this case $U(x)$ gives the full single-particle
potential shape in the QW. If $n_b>n_0$ (this corresponds to $\Delta\mu>0$),
Fig.~\ref{n-b15} shows that the potential shape lies below the $U_0$ energy
and hence the interaction energy is negative.  This means that a positive
charge is accumulated in the QW. It is responsible for the appearance of a
potential well.

The case $\Delta\mu<0$ is realized when $n_b<n_0$. The electron density
distribution and the potential shape are shown in Fig.~\ref{n-b05}. The
interaction energy is seen to be positive and the potential shape in the QW
lies above $U_0$. This means that a negative charge is accumulated in the QW.
It produces a potential barrier that hinders electrons in passing through
the QW.

If $n_b= n_0$, the electron density is essentially not redistributed
between the QW and the reservoirs. However, Friedel oscillations of the
electron density appear near the contacts, Fig.~\ref{n-b10}. The Friedel
oscillations are also observed if there is carrier transfer, but they are
superimposed on the much stronger variation of the potential due to the
charging of the QW.

Let us now consider the exchange interaction effect. The electron density
distribution and the potential shape, calculated by taking into account the
exchange interaction, are shown in Figs.~\ref{n-b15}, \ref{n-b05},
\ref{n-b10} as solid lines.  The exchange interaction is seen to result in
an increase of the electron density. This is a consequence of the fact
mentioned in Sec.~\ref{Contact-potential} that the exchange interaction
decreases the chemical potential in a QW. That is why more electrons come
into the QW when the equilibrium state is established.  Correspondingly, the
negative charge in the QW increases, which results in the growth of the
energy $U(x)$. However, this does not yet mean that the electron states with
energy lower than $U(x)$ are necessarily states decaying in the QW,
since the exchange interaction lowers the effective barrier between
the QW and the reservoir.

\begin{figure}[htb]
\mbox{\epsfig{file=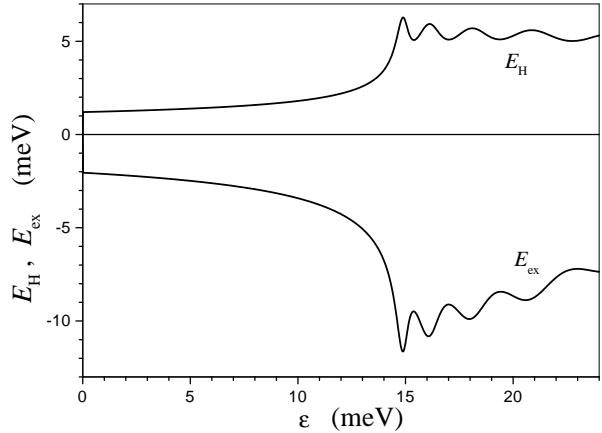,width=9cm}}
\narrowtext
\caption[to]{Average exchange and Hartree energies of electrons incident on
the QW with the energy $\varepsilon$. The calculations are for the
parameters:  $a= 5\times 10^{-7}$~cm, $L/a = 30$, $U_0$ = 20~meV, $\mu_0-U_0 =
4$~meV, $n_b= 1.0n_0$, $T$= 0, and $V_a$= 0.}
\label{Ex-EH}
\end{figure}

Since it would be incorrect to consider the exchange energy as a function of
$x$, we calculate an average value of the exchange energy per particle
incident on the QW with the energy $\varepsilon(k_+)$ from the left
reservoir,
$$
E_{\rm ex} = \frac{\langle \psi |\hat{H}_{\rm ex}|\psi \rangle}
{\langle \psi |\psi \rangle}\,,
$$
(here $\langle\cdots \rangle$ denotes averaging over the QW length). The
average exchange energy $E_{\rm ex}$ is shown in Fig.~\ref{Ex-EH} as a
function of $\varepsilon(k_+)$. The average Hartree energy $E_H$ is also
given in this figure. It is seen that $E_{\rm ex}$ exceeds $E_H$ for all
energies. Hence the joint effect of the exchange interaction and the Hartree
interaction consists in an effective lowering of the barrier, so that
electrons with energy below the confinement energy $U_0$ can transit through
the QW without decay.

Another effect produced by the exchange interaction is also seen from
Figs.~\ref{n-b15}, \ref{n-b05}, \ref{n-b10}. The exchange interaction strongly
enhances the Friedel oscillation amplitude. This result agrees qualitatively
with the analytical calculation of the interaction effect on the transmission
through a barrier in 1D systems.~\cite{Yue} In our case the Friedel
oscillations are generated at the contacts of the QW with the reservoirs.
Their amplitude in the potential energy is quite pronounced but smaller than
the Fermi energy $\mu_0-U_0$ in the QW. For the discussion which follows it
is important to remark that the Friedel oscillations are superimposed on the
smooth variation of the potential produced by the charge accumulated in the
QW. Even if the exchange interaction is fully taken into account, this smooth
component has an amplitude that is larger than that of the Friedel
oscillations.  Due to the smooth variation of the potential the QW becomes
nonuniform.

\section{Linear conductance}
\begin{figure}[htb]
\mbox{\epsfig{file= 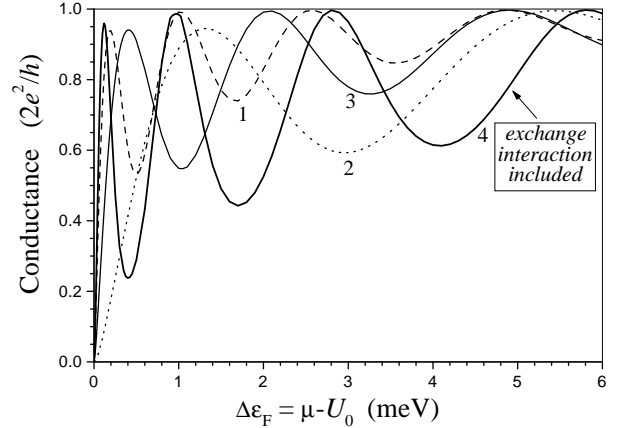,width= 9cm}}
\narrowtext
\caption[to]{Linear conductance as a function of the chemical potential
$\mu_0$.  Lines 1, 2, and 3 present the calculations within the Hartree
approximation for several background densities: $n_b/n_0$ = 1.5, 0.5, 1.0,
respectively. Line 4 is obtained by including the exchange interaction for
$n_b= n_0$. The parameters used in the calculation are $a= 5\times
10^{-7}$~cm, $L/a= 20$, $U_0$= 10~meV, $T$= 0, and $V_a$= 0.01~mV.}
\label{L-cond}
\end{figure}

The model that we have developed above allows us to find the electric
current arising when an external voltage is applied.  The current is
calculated as the sum of the partial currents of the states $\psi_{r,k_r,s}$,
taking into account their occupation. The linear regime is realized when
$eV_a \ll (\mu_0 - U_0)$. In this case we have obtained the dc conductance as
a function of the chemical potential $\mu_0$. The results of these
calculations for zero temperature and for several densities of the background
charge $n_b$ are given in Fig.~\ref{L-cond}.

The conductance oscillations with varying chemical potential are
a consequence of the nonadiabatic reservoir-wire interface. The rapid
variation of the potential at this interface leads to backscattering and, if
the electron wave is coherent over the entire wire length, to resonances. The
oscillations have the same origin as the resonances observed in over the
barrier transmission of noninteracting particles.  A similar effect also
appears in the transport of noninteracting electrons through a narrow,
ballistic, nonadiabatic constriction in a 2D electron gas.~\cite{Kirczenow}
Our calculations show that the {\em e-e} interaction changes the effective
potential barrier that electrons have to overcome in passing from one reservoir
to another.

First, we discuss the results obtained within the Hartree approximation, when
exchange interaction is neglected. These results are represented by the lines
1-3 in Fig.~\ref{L-cond} for several different background densities $n_b$.
With increasing $n_b$ the oscillations become more frequent, which means that
the effective wave number of the electrons is increased. Exactly the same
behavior is demonstrated in Figs.~\ref{n-b15}, \ref{n-b05}, \ref{n-b10},
namely, with increasing $n_b$ the potential in the QW is shifted downward
causing the kinetic energy to increase.

\begin{figure}[htb]
\mbox{\epsfig{file= 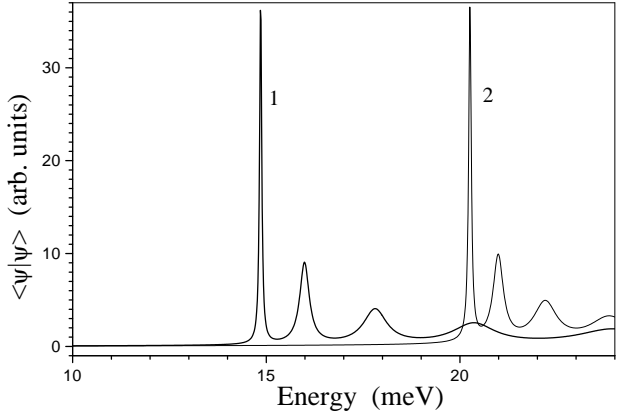,width= 9cm}}
\narrowtext
\caption[to]{Exchange interaction effect on the spectrum of the electron
density
$\langle \psi_k|\psi_k\rangle$. Line 1 presents the spectral density
obtained by including the exchange interaction; line 2 is the spectral
density without the exchange interaction. The calculations are for:  $a=
5\times 10^{-7}$~cm, $L/a= 30$, $U_0$= 20~meV, $\mu_0-U_0= 4$~meV, $n_b=
1.0n_0$, $T$= 0, and $V_a$= 0.}
\label{psi}
\end{figure}

A similar effect occurs when the exchange interaction is turned on. It is
demonstrated in Fig.~\ref{L-cond} by curve 3 (obtained by ignoring the
exchange interaction) and by line 4 (obtained by including
exchange interaction), the
background charge being the same in both cases. The exchange interaction is
seen to make the conductance oscillations more frequent. The reason for this
effect is that the exchange interaction results in an effective lowering of
the potential energy of the electrons and consequently in an increase of their
kinetic energy. In order to assess the exchange interaction effect on the
effective potential, it is instructive to see how the exchange interaction,
affects the spectral density of electrons, i. e., $|\psi_k|^2$
integrated over the QW length. This is illustrated in Fig.~\ref{psi}.  The
exchange interaction allows the electrons with energy below the confinement
energy $U_0$ to pass through the QW.

It is interesting to note that, despite the fact that Friedel oscillations
are present in the QW, no noticeable suppression of the conductance is
observed.  The effect of conductance suppression by a periodic potential
associated with Friedel oscillations was considered for infinite 1D
systems with a $\delta$ potential in Refs~\onlinecite{Kane} and
\onlinecite{Yue}.  This phenomenon is connected with the fact that a periodic
component of the potential suppresses the transmission of the electrons with
energy near the Fermi level across the QW (a gap appears at the Fermi level).
The absence of this effect in our system is a consequence of two facts.
First, the QW has a finite length. Second (and no less essential), the QW becomes
inhomogeneous owing to the electron density redistribution between the QW and
the leads. As a consequence the kinetic energy at the chemical potential
level and the Friedel oscillation period become dependent on the position in
the QW. This is why the resonant interaction of electrons at the
chemical potential level with the Friedel oscillations is destroyed and the
electron passage is not suppressed.

\section{Nonlinear transport and multistability}
\label{multi-stability}
A significant redistribution of the electron density between the QW and
the reservoirs occurs under far from equilibrium conditions when the applied
voltage exceeds the Fermi energy. Electrons are injected from the left
reservoir (cathode) while the electrons entering the QW from the positive
reservoir (anode) are scattered back inside the QW. As a consequence, the
electron density decreases in the QW (roughly speaking to one-half of the
equilibrium density) though the positive background charge is unchanged.
Because the positive charge is dominant, a potential well appears in the QW,
with the potential shape being distorted by the external potential, as
illustrated in Fig.~\ref{diagram}. Therefore the kinetic and potential
energies are greatly changed. The change in the potential energy produces
variations in the wave functions (including even a possibility for resonant
states to appear) and the electron density distribution. In this way
feedback arises between the electron density and the potential in the QW,
which is an important mechanism in nonlinear transport. It is that mechanism
which is realized in the model proposed.

A complete numerical analysis of the nonlinear transport properties in the
wide range of applied voltages within the Hartree-Fock approximation meets
some difficulties caused by the long computation time. In this paper we
restrict our consideration to the Hartree approximation, which is reasonable
at high enough voltage because the exchange energy decreases when the kinetic
energy of electrons is increased.~\cite{Mahan} The calculations were carried
out using the method of pseudotime evolution to the steady
solution~\cite{Samarskii} described in the Appendix. It turns out that in
some range of applied voltage an instability of the evolution process
appears.  The instability origin is not connected with the computation
process but is caused by real behavior of the system.

The mechanism of the instability is as follows. When the applied voltage
is high enough (compared to the Fermi energy), the electron flow injected
from the negatively charged reservoir is the only flow in the QW. Let a
velocity fluctuation appear in some portion of the wire. To be definite, let
us assume that the velocity is increased above its stationary value. Since
the total electron flow is limited by the contact, it is not disturbed by
this fluctuation.  Hence, the continuity of the current requires that the
electron density decreases. This leads to a growth of the positive (net)
charge, because electrons cannot completely neutralize the background
charge. The excess positive charge causes the potential energy of the
electrons to decrease. Under the condition of ballistic transport, this
results in a new increase of the velocity, and so on until some nonlinear
process stabilizes this instability. In our model this is achieved by a
redistribution of the overall electron density and a reshaping of the
potential distribution in the QW.  In such a way the potential shape is
switched from one state to the other under the condition that both states are
characterized by the same potential difference across the QW ends.

\begin{figure}[htb]
\mbox{\epsfig{file=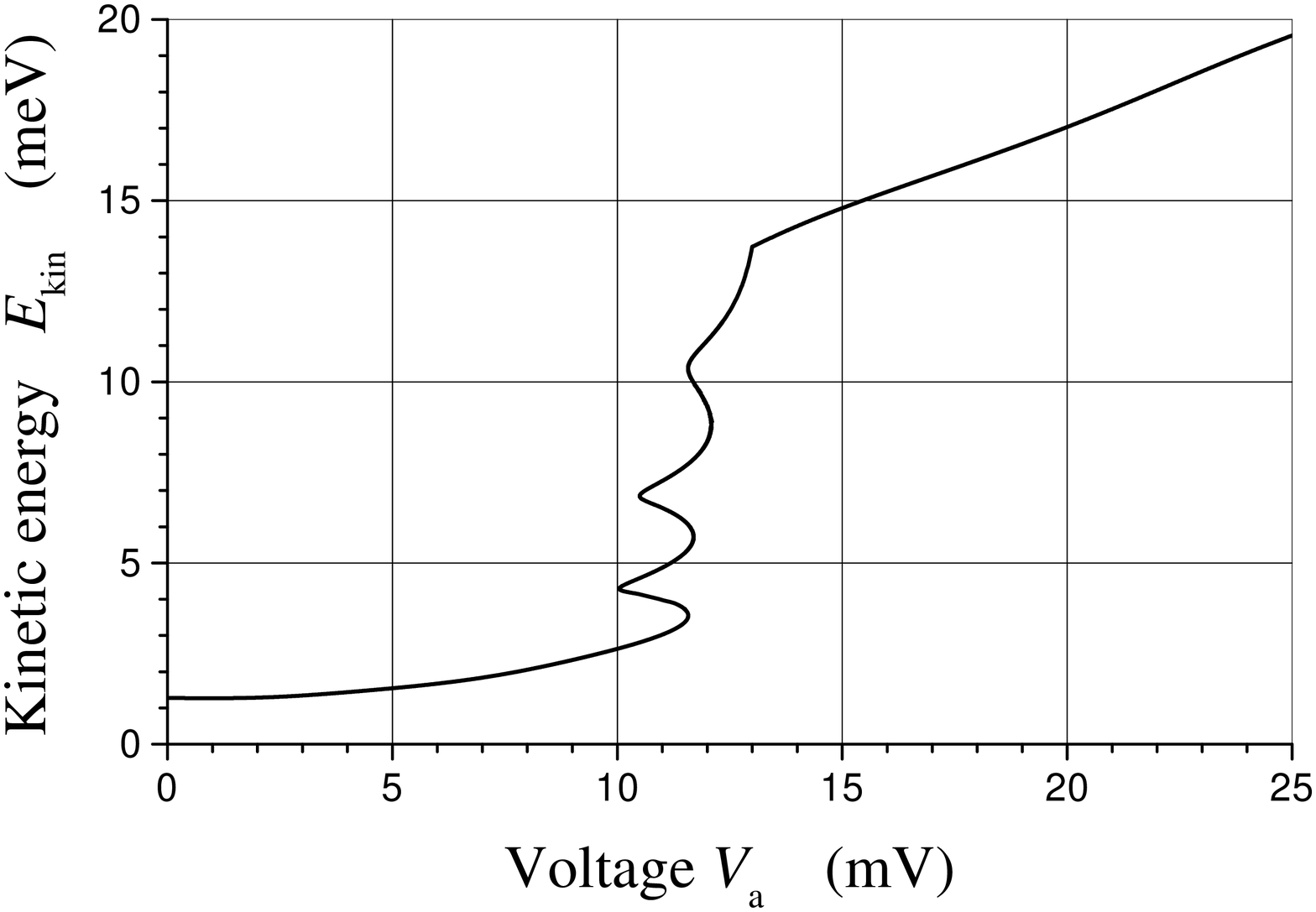,width=9cm}}
\narrowtext
\caption[to]{Dependence of the average kinetic energy of electrons on the
applied voltage. The calculations are for the following parameters:
$a= 5\times 10^{-7}$~cm, $L/a= 20$, $U_0$= 10~meV, $\mu_0-U_0= 3$~meV,
$n_b/n_0$= 1.0, and $T$= 0.}
\label{multi-1}
\end{figure}

In order to describe the transition from one shape to the other as a
continuous process it is necessarily to characterize the nonequilibrium state
of the system by a parameter other than the applied voltage. This parameter
should distinguish states with different potential shapes and the same
potential difference across the ends. As such a parameter, the mean kinetic
energy $E_{\rm kin}$ of electrons in the QW can be used,
$$
E_{\rm kin} =
\frac{\sum_{r=\pm} \int dk_r \langle k_r |{\hat T}|k_r \rangle}
{\int_{-L/2}^{L/2}dx\,n(x)}\;,
$$
where $\hat T$ is the kinetic energy operator.

This conclusion is similar to what is known in the theory of hot electron
instabilities in semiconductors. The electron heating by the electric field
results in S- and N-shaped current-voltage characteristics. Under these
conditions, the electron temperature uniquely determines the state of the
system.~\cite{Bonch} In our case the kinetic energy of the electrons is a
direct analog of the electron temperature.

We have developed the algorithm that allows one to solve our problem in the
case where the mean kinetic energy is fixed rather than the applied voltage.
This algorithm is described in the Appendix. In this computation scheme all
quantities (including $V_a$) are determined by $E_{\rm kin}$. We have found
that this algorithm gives stable results and the nonequilibrium state of
the system under investigation is uniquely determined by $E_{\rm kin}$. In
particular, the dependence of $E_{\rm kin}$ on the applied voltage is shown
in Fig.~\ref{multi-1}. The kinetic energy is seen to have several values for
a given voltage $V_a$, while $V_a$ is uniquely defined by $E_{\rm kin}$.
Correspondingly, several nonequilibrium states, with different distributions
of the electron density and the potential, are possible at a given voltage.
It is obvious that the states are not all stable with respect to
time-dependent fluctuations.

The multistability, and in particular bistability, phenomenon described
above may be useful for understanding the negative differential conductance
observed in quantum wires.~\cite{Berven}

\section{Conclusions}
In this paper a QW of a finite length coupled to reservoirs is considered as
a unified system. We have found that the electron density is substantially
redistributed between the wire and the electron reservoirs when this system
is formed. As a consequence of this process, a QW can acquire a net charge.
The charging of the wire is caused by the chemical potential difference
between the wire and the leads which exists if these two subsystems are
decoupled.  This phenomenon is similar to the contact potential difference in
a classical multiconductor system.  The structures of the charge density
and the electric potential distributions differ according to the chemical
potential difference:  (i) a positive charge is accumulated in the wire
and a potential well is developed there; (ii) the wire is charged
negatively giving rise to a potential barrier; (iii) the wire remains
uncharged as a whole.  In all cases, Friedel oscillations are present which
are generated at the nonadiabatic contacts of the QW with reservoirs.  They
are superimposed on the relatively smooth profile of the potential produced
by the charge accumulated in the wire. This smooth potential has a large
amplitude.
The Friedel oscillation amplitude is strongly enhanced if the exchange
interaction is included.

Variation of the electron density in a QW due to electron redistribution
between the wire and the reservoirs produces a significant effect on dc
conductance. This effect is connected with the change of the kinetic energy
of the electrons due to two factors:
(i) the variation of the accumulated charge and the potential
variation associated with this charge;
(ii) the variation of the exchange energy. The exchange energy
substantially lowers the effective potential barrier that electrons have to
overcome when passing from one reservoir to another.

The Friedel oscillation potential does not suppress the conductance because
the QW becomes inhomogeneous as a consequence of the electron density
redistribution between the QW and the leads.
The electron density variation due to the voltage applied across the leads
and the change of the effective potential shape, associated with this
variation, in the wire is an important mechanism for nonlinear transport.

The most interesting consequence of the charging effect in the wire is the
instability that arises under a high enough applied voltage. The instability
shows itself as a spontaneous increase of the kinetic energy of the injected
electrons at a given applied voltage. In a QW with a nonadiabatic connection
to the reservoirs, the development of instabilities results in the appearance
of multistable states, i.e., in the existence of several stable states
at a given voltage.

We conclude by emphasizing that the charging effect analyzed here is a
general phenomenon that might be important in many mesoscopic systems
containing electronically different compounds. Examples  of current interest
are carbon nanotubes,~\cite{Datta} hybrid normal-superconducting systems, and
atomic quantum point contacts.

\section*{Acknowledgments}
The present work has been supported by INTAS (Grant No. 96-0721).
V.A.S. and S.V.P acknowledge the Russian Fund for Basic Research (Grant No.
99-02-18192), the Russian Program "Physics of Solid-State Nanostructures"
(Grant No. 97-1054) and the Russian Program "Surface Atomic Structures" (Grant
No. 5.3.99). M.B. is funded by the Swiss National Science Foundation.
The multiprocessor computer system Parsytec CC used for the numerical
calculations has been acquired by the Institute for Mathematical Modeling of
the Russian Academy of Sciences with financial support from the EC ESPRIT
program (Grant No. 21041).

\appendix

\renewcommand {\theequation}{\thesection.\arabic{equation}}
\section{The computation method}
\setcounter{equation}{0}

The problem of finding the wave functions for a QW coupled to reservoirs can
be reduced to the solution of Eq.~(\ref{Schrodinger}) in the inner region
$-L/2 \le x \le L/2$ and matching $\psi_{r,k,s}$ and its derivative at the
boundaries $x=\pm L/2$ to the wave functions (\ref{reservoirWavefunction}) in
the outer regions. Combining the matching equations for $\psi_{r,k,s}$ and
$\psi'_{r,k,s}$, one can exclude the coefficients $R_r$ and $T_r$ to get
finally the following nonuniform boundary conditions for $\psi_{r,k,s}$:
\begin{equation}
\label{boundary_cond}
\psi'_{r,k_r,s} = \left\{
\begin{array}{lcl}
i k_r (2-\psi_{r,k_r,s})\,,\quad & & rx=-L/2\\
i k'_r\psi_{r,k_r,s}\,,\quad & & rx=L/2\,,
\end{array}
\right.\;,
\end{equation}
where $k_r$ and $k'_r$ are defined in Eq.~(\ref{reservoirWavefunction})

The wave functions $\psi_{r,k_r,s}$ are considered as functions of two
continuous variables: the space coordinate $x$ and the energy
$\varepsilon_r=\hbar^2k_r^2/2m$. Thus
$\psi_{r,k_r,s}=\psi_{r,s}(x,\varepsilon_r)$, where $x$ and $\varepsilon_r$
are varied respectively in the regions $-L/2 \le x \le L/2$ and
$0<\varepsilon_r \le\varepsilon_M$, with the upper boundary $\varepsilon_M$
being equal to $\mu_0+3k_BT$ ($k_B$ is the Bolzmann constant, $T$ is the
temperature). In this region the uniform grid
$\{x_i,\varepsilon_j\}_{i=0,N_1; j=0,N_2}$ is constructed.

The equation for $\psi_{r,k,s}(x_i,\varepsilon_j)$ on the grid is obtained
from Eq.~(\ref{Schrodinger}) with use of the integro-interpolative
method for the node presentation on $x_i$ and the trapezium formula when
calculating the integrals on $k_r$ for the nodes on $\varepsilon_j$. This
results in a set of nonlinear finite-difference equations that can be
symbolically presented in the form
\begin{equation}
\label{M_Psi}
{\rm \bf \hat{M}[\Psi] \Psi} = {\rm \bf F}\;,
\end{equation}
where ${\bf \Psi}$ is the wave function vector to be found and ${\rm \bf
\hat{M}[\Psi]}$ is the nonlinear operator. The matrix equation~(\ref{M_Psi})
is inhomogeneous as a consequence of the boundary
conditions~(\ref{boundary_cond}).

The equations~(\ref{M_Psi}) are solved by the iteration method. However, the
commonly used successive approximation scheme (${\rm \bf
\hat{M}}[{\bf \Psi}^{(l)}] {\bf \Psi}^{(l+1)}={\rm \bf F}$, with $l$ being
the iteration number) turns out to be badly convergent. We use the method of
pseudotime evolution to the steady solution.~\cite{Samarskii}
More specifically, we use the two-layer iteration scheme of this method. In
this scheme the approximating matrix ${\rm \bf \hat{M}}^{(l)}$ is introduced,
which is calculated with the use of the iteration process
\begin{equation}
\label{M-evolution}
\frac{{\rm\bf \hat{M}}^{(l+1)}-{\rm\bf \hat{M}}^{(l)}}{\tau_l}=
{\rm\bf \hat{M}}[\Psi^{(l)}]-{\rm\bf \hat{M}}^{(l)},\;
l = 0, 1, 2, \dots\,,
\end{equation}
where $\tau_l$ is a pseudotime parameter. The choice
of $\tau_l$ allows one to attain the best convergence of the iteration
process. As the starting value of ${\rm \bf \hat{M}}^{(0)}$ we use ${\rm \bf
\hat{M}}[\Psi=0]$, i.e., the ${\rm \bf \hat{M}}$ matrix for
noninteracting electrons. During the iteration process, $\Psi^{(l)}$ is
calculated with the use of the equation
$$
{\rm\bf \hat{M}}^{(l)}{\rm\bf\Psi}^{(l)}={\rm\bf F}\;.
$$
The pseudotime $\tau_l$ is determined by the ${\rm\bf \hat{M}}^{(l)}$
operator spectrum. The optimal convergence is attained when
$$
\tau_l=\frac{2}{\lambda_{\rm min}(M^{(l)})+\lambda_{\rm max}(M^{(l)})}\;,
$$
where $\lambda_{\rm min}$ and $\lambda_{\rm max}$ are the lowest and highest
eigenvalues of ${\rm\bf \hat{M}}^{(l)}$.
The iteration process is ended when the following condition is fulfilled
$$
{\rm max} \left|\frac{m^{(l+1)}_{ij}- m^{(l)}_{ij}}{ m^{(l)}_{ij}}\right| \le
\delta\;,
$$
where $m^{(l)}_{ij}$ is an element of the ${\rm\bf M}^{(l)}$ matrix. In the
present paper, $\delta$ was chosen to be $10^{-6}$.

The above method is successful when the system under investigation has a
unique solution. However, at some fixed values of the applied voltage the
computation shows an instability. In the course of the pseudotime
evolution process the calculated quantities (such as the potential, the
electron density, the kinetic energy) are randomly switched between several
values.  This is connected with the fact that the state of the system is not
uniquely determined by the calculation scheme where the applied voltage is
fixed.

A unique description of the system is achieved by using the mean kinetic
energy $E_{\rm kin}$ of the electrons as the parameter that defines the
nonequilibrium state of the system. We have developed computation
algorithm that allows one to vary $E_{\rm kin}$ continuously, in other
words, we solve the problem using $E_{\rm kin}$ as the fixed parameter
instead of the applied voltage.

An essential question appearing in this algorithm is how the applied voltage
$V_a$ should be defined when $E_{\rm kin}$ is given. The equation defining
$V_a$ is obtained from Eq.~(\ref{Schrodinger}). Multiplying this equation by
$\psi^*_{r,k_r,s}$, integrating it over $k_r$ and over $x$, and summing over
$r$ one gets an equation of the following form:
\begin{equation}
E_{\rm kin} - eV_a A[{\bf \Psi}] = B[{\bf \Psi}]\,,
\end{equation}
where $A[{\bf \Psi}]$ and $B[{\bf \Psi}]$ are functionals of the electron
wave functions. Solving this equation with respect to $V_a$ one gets $V_a$ as
a functional of ${\bf \Psi}$, with $E_{\rm kin}$ being a parameter,
\begin{equation}
\label{V_a-psi}
V_a=\Phi_{E_{\rm kin}}[{\bf \Psi}]\;.
\end{equation}

When solving the problem with $E_{\rm kin}$ as a parameter,
Eq.~(\ref{V_a-psi}) should be taken into account together with
Eq.~(\ref{M-evolution}). This system of equations is solved using the above
pseudotime evolution method and two-layer iteration scheme. The set
$V_a^{(l)}$ approximating $V_a$ is defined as
$$
\frac{V_a^{(l+1)}-V_a^{(l)}}{\tau_l} =
\Phi_{E_{\rm kin}}[{\bf\Psi}^{(l)}] - V_a^{(l)}\;,\quad l = 0, 1, 2, \dots\;.
$$
As the starting value of the $V_a^{(l)}$ set, the arbitrary value of $V_a$ in
the stability region close to the instability threshold can be used. In this
generalized procedure the pseudotime $\tau_l$ is chosen taking into account
the spectral properties of the total matrix ${\rm\bf M}^{(l)}\oplus
V_a^{(l)}$.

\end{multicols}

\begin{thebibliography}{40}
\bibitem{footnote1} It is necessary to bear in mind that, for a wire
of finite length, the chemical potential depends on the wire length.
\bibitem{footnote2} Our analysis shows that localized states may appear
far from equilibrium, if the background charge density is
much higher than $en_0$ defined in Eq.~(\ref{n_0}). This case should be
investigated separately.
\end{thebibliography}
\end{document}